\documentclass[aps,prb,preprint,endfloats]{revtex4}
\makeatletter
\def\@dotsep{4.5}
\makeatother
\usepackage{graphicx}
\begin{document}

\title{A novel method for evaluating the critical nucleus and the surface tension in systems with first order phase transition}

\author{Chiara Cammarota}
 \email{Chiara.Cammarota@roma1.infn.it}
\affiliation{%
Dipartimento di Fisica, 'La Sapienza' Universit\`a degli studi di Roma,
Piazzale Aldo Moro 5, 00185 Roma, Italy and \\
Center for Statistical Mechanics and Complexity, (SMC), CNR - INFM, via dei Taurini 19, 00185 Roma, Italy
}

\author{Andrea Cavagna}
\affiliation{
Center for Statistical Mechanics and Complexity (SMC), CNR -INFM, via dei Taurini 19, 00185 Roma, Italy
and Istituto Sistemi Complessi (ISC), CNR, via dei Taurini 19, 00185 Roma, Italy 
}

\date{\today}

\begin{abstract}
We introduce a novel method for calculating the size of the critical nucleus
and the value of the surface tension in systems with first order phase transition.
The method is based on classical nucleation theory, and it 
consists in studying the thermodynamics of a sphere of given radius
embedded in a frozen metastable surrounding. The frozen configuration
creates a pinning field on the surface of the free sphere. The pinning field forces the
sphere to stay in the metastable phase as long as its size is smaller
than the critical nucleus. We test our method in two first-order systems, both
on a two-dimensional lattice: a system where the parameter tuning the
transition is the magnetic field, and a second system where the
tuning parameter is the temperature. In both cases the results are satisfying.
Unlike previous techniques, our method does not 
require an infinite volume limit to compute the surface 
tension, and it therefore gives reliable estimates even by using relatively 
small systems. However, our method cannot be used at, or close to, the critical point, i.e. 
at coexistence, where the critical nucleus becomes infinitely large.

\end{abstract}

\maketitle

\section{\label{introduction} Introduction}

Whenever in a system there is coexistence of two (or more)
thermodynamic phases, the crucial issue arises of how to compute the
free energy cost associated to creating a surface among the different
phases.  This cost is the {\it surface tension}~\cite{binder87}.  The surface tension
is relevant in the case of broken ergodicity at equilibrium, when the
two phases have the same free energy density, as for example in the
Ising model below $T_c$ at zero magnetic field: equilibrium droplet
excitations of a phase within the dominating opposite phase, as well
as off-equilibrium coarsening dynamics of competing domains, are all
phenomena quantitatively ruled by the surface tension cost. The
surface tension is in fact related to the prefactor of the square
gradient term suppressing fluctuations of the order parameter in any
field-theoretical approach to statistical physics.

Surface tension is also a key ingredient in nucleation theory, that is
the theory of how metastable states decay when crossing a first order
critical point.  In this case the two phases do not have the same
free-energy density. Rather, the metastable phase has higher free
energy, and it decays by nucleating droplets of the stable phase. The
thermodynamic advantage of a nucleus is proportional to its volume,
while its cost is proportional to the surface. Therefore, nucleation
is a competition between the bulk free energy difference and the
surface tension between the two phases.  As a consequence, computing
the surface tension is essential also in this context.

Numerical determinations of the surface tension are not trivial.
Leamy and coworkers~\cite{1973PhRvL..30..601L} compared two systems:
one in which opposite boundary conditions give origin to an interface
between two phases with the same free-energy, and a second one with
uniform boundary conditions.  They numerically evaluated the extra
free-energy density profile of a dividing surface (essentially flat up
to fluctuations) and the density variation across this surface.  Some
years later Binder~\cite{1982PhRvA..25.1699B} presented a method
suitable for studying the free-energy excess density near the critical
point, where the correlation length is large and the excess
free-energy is small and strongly fluctuating. Binder's method
therefore works well in a region where the Leamy method is not
accurate. The main idea of this method is that in a broken symmetry phase,
the (small) probability of having order parameter equal to zero is 
dominated by configurations with phase separation, where the surface tension
contribution is the most relevant. Binder's method, currently the
standard for computing the surface tension, is theoretically very well
motivated and it shows a good agreement with experimental
data~\cite{sp1, sp2, sp3, sp4}, even compared with the results of some
analytical approaches~\cite{ch1, fiskwidom, OK}. However, this method
requires the introduction of several fitting parameters, and, more
importantly, the results depend on the infinite-size extrapolation.

Both these methods compute the surface tension between two phases of 
equal free-energy density. However in the context of first-order transition 
and nucleation theory, we deal with the free-energy cost of surfaces between 
a stable and a metastable phase. The theoretical context is therefore slightly
different, as are the numerical methods to compute the surface tension.
The transition path sampling (TPS) 
method~\cite{1998JChPh.108.1964D, 1998JChPh.108.9236D, chandler, 1999JChPh.110.6617D, ising, neutronscattering}, 
is based on the idea of sampling the distribution of paths in the phase space 
with a Monte-Carlo procedure analogous to the sampling of the canonical probability 
density of many-particle system.
The TPS method can be used in the context of nucleation, but more generally it 
is a very  useful tool for all dynamical problems dominated by rare events, 
as the dissociation of weak acid in water or protein folding. 
Due to its intrinsic generality, the TPS method is numerically heavier and somewhat  
less suited to the study of first order transitions, than a method directly inspired 
by nucleation theory. The aim of the present work is to introduce such a method.

The main idea of our method is to study the thermodynamics a sphere of
radius $R$ embedded in a system frozen into a typical metastable
configuration. The metastable surrounding produces a frozen pinning
field (FPF) at the surface of the sphere, which biases the Gibbs
distribution of the sphere. For small $R$ the sphere is kept in the
metastable phase by the pinning field, whereas for large $R$ the
free-energy gap drives the sphere to the stable phase. The value of
$R$ where the probability for the sphere to be in the metastable phase
is the same as in the stable one gives the critical nucleus, and from
this the surface tension can easily be calculated.  The FPF idea of
using constrained systems to modify the Gibbs measure and magnify the
effect of metastability, is well known in disordered systems, and in
particular in spin-glasses~\cite{kurchan93,franz95,monasson95,cavagna97}.  
In fact, the FPF method was originally
proposed for the study of glassy systems to test the Mosaic
scenario~\cite{2004JChPh.121.7347B,juanpe,2006cond.mat..7817C}.

In Sec. \ref{themethod} we will make a short review of classic
nucleation theory and explain the FPF method in general. As a warm-up
exercise, in Sec. \ref{t>tc} we will study the method in absence of
metastability, where it will turn out to be a tool for computing the
correlation length of the system.  In Sec. \ref{critclsize} the FPF
method is applied to the ferromagnetic Ising model $d=2$, with nonzero
magnetic field. In this case there is a first-order phase transition
driven by the field, and we will compute the surface tension at
different values of the magnetic field and of the temperature.  The
obtained surface tension values will be compared with the exact
Onsager results. In Sec. \ref{ctls} we will consider a different
first-order model, where the tuning parameter is temperature, rather
than field. This case is interesting, since it mimics the
liquid-crystal transition, and it allows to investigate the method in
the case of a disorder-order phase transition.

\section{\label{themethod} The Frozen Pinning Field method}

The frozen pinning field (FPF) method is inspired by homogeneous
nucleation theory, originally introduced by Gibbs in 1948~\cite{gibbs}. 
A system in a metastable phase makes the transition via
the formation of droplets (nuclei) of the stable phase.  The
free-energy cost, $\Delta F(R)$, of a nucleus of linear size $R$ is
given by the balance between a volume gain, proportional to the
bulk free-energy density difference between the two phases $\delta f$,
and a surface price, proportional to the surface tension $\sigma$,
\begin{equation}
\Delta F(R)=-V_d R^d\delta f +S_d R^{d-1}\sigma  \ ,
\label{DF}
\end{equation}
where $V_d$ and $S_d$ are the volume and surface numerical prefactors related 
to the shape of the droplet in dimension $d$. The critical nucleus size, 
\begin{equation}
\widehat R=\frac{(d-1)\, S_d}{d\, V_d}\ \frac{\sigma}{\delta f}\:,
\label{Rmax}
\end{equation}
that maximizes Eq. (\ref{DF}), distinguishes smaller unstable nuclei that will shrink, 
from bigger stable nuclei, which are thermodynamically favoured to grow. When  
the volume gain exceeds the surface cost, the nuclei become stable, and the system
is heterogeneous. At that point growth can start, and it will eventually drive the 
whole system to the stable phase.

The calculation of $\widehat R$, and therefore of $\sigma$, is 
typically done by calculating the time needed to form a critical
nucleus, i.e. the {\it nucleation time}. 
From (\ref{DF})
we have that the barrier to nucleation is given by,
\begin{equation}
\Delta F(\widehat R)= A \frac{\sigma^{d-1}}{\delta f^d}
\end{equation}
where $A$ contains all geometrical factor. The nucleation time is simply the 
time needed to cross this barrier, which, according to Arrhenius formula,
is given by,
\begin{equation}
\tau_N=\tau_0\; \exp\left( \beta A \frac{\sigma^{d-1}}{\delta f^d}\right) \ .
\label{tempo}
\end{equation}
In finite dimension this time is infinite only at the transition
point, where $\delta f=0$, and at zero temperature, where
$\beta=\infty$. Therefore, in finite dimension the decay time of a
metastable state is always finite. For this reason, of course, giving
a sharp theoretical definition of a metastable state in finite
dimension is very difficult~\cite{La1, La2,binder87}. 
In our work, we will simply give an operational definition of
metastability.  This can be done by comparing the nucleation time
$\tau_N$ with the equilibration time of the metastable phase, or
relaxation time, $\tau_R$. As long as,
\begin{equation}
\tau_N \gg \tau_R \ ,
\label{tomarchio}
\end{equation}
the metastable phase is operationally well defined, in the sense that
it is possible to measure observables which are in local equilibrium,
well before the metastable phase decays and becomes heterogeneous~\cite{binder87}. 
By varying the external parameters, one may reach the {\it
kinetic spinodal} point, defined by the relation,
\begin{equation}
\tau_N \sim \tau_R \quad\quad \quad \mathrm{kinetic\ \ spinodal}   \ .
\end{equation}
Beyond this point the metastable phase is no longer well defined,
since the time needed to relax any perturbation is of the same order
as the time needed to form critical nuclei of the stable phase. In
what follows we will always operate under the condition of equation
(\ref{tomarchio}).  In the cases analyzed in the present paper the
parameter tuning the first-order transition will be either the
magnetic field $h$, or the temperature $T$. However, to fix ideas in
the general discussion we will consider a 'thermal' setup, with a
transition temperature $T_c$ (where $\delta f=0$), below which one of
the two phases becomes metastable, and a kinetic spinodal temperature
$T_{sp}$, below which the metastable phase becomes kinetically
unstable. All our discussion will thus be staged for $T_{sp} < T <
T_c$.  To avoid misunderstandings, we also note that throughout the
paper we will consider a metastable phase to be in local equilibrium
if one-time quantities (as the energy) do not depend on time, and if
two-time quantities (as the correlation function) depend only on the
difference of times (which ensures that the fluctuation dissipation
theorem holds).

One could think that by measuring the nucleation time $\tau_N$, it is
possible to directly access the surface tension $\sigma$, provided that
the $\delta f$ is known. However, this is practically very hard to do,
and for many good reasons. First, determining the prefactor $\tau_0$
in (\ref{tempo}) is far from trivial \cite{turnbull}. Nucleation is a
dynamical aggregation process, and for this reason it is well known
that $\tau_0$ is {\it not} a constant. For example, in liquids it will
strongly depend on the temperature, possibly through the diffusion
constant $D$.  However, it is still not well established the exact
form of the dependence of $\tau_0$ on the external parameters, like
temperature, density, field, etc. Moreover, the result of classic
nucleation theory is not the nucleation time, but rather the
nucleation rate {\it per unit volume}.  This is obvious: larger
systems have a larger probability to produce stable nuclei
\cite{nucleazione}. The consequence is that the prefactor $\tau_0$
contains also a volume factor, which may not be easy to determine.
Secondly, the geometric factors $V_d$ and $S_d$, which enter in $A$,
are not known for general spontaneous droplets, which may have any
sort of complicated shape. Note that these factors enter in the
exponential, so they are not at all irrelevant. Third, the very
calculation of the nucleation time is very difficult, and most of the
times it is done inaccurately. A typical protocol is to fix the
nucleation time as the time when the amount of stable phase (for
example, the crystal) in the system reaches a certain arbitrary
threshold.  The problem is that by doing this one is mixing {\it
nucleation} and {\it growth}, which, however, are different
phenomena~\cite{2003JChPh.118.6974C}. 
In fact, there may be cases
when nucleation is very fast, but growth is exceedingly slow, due to
the very small mobility of the system. Measuring a {\it bona fide}
nucleation time is extremely difficult, since it requires knowing the
size of the critical nucleus.  But if we knew it, we would have no
need to find out the nucleation time to access the surface tension!

The FPF method avoids these problems, giving us direct access to the critical 
nucleus. The method can be used both in off-lattice and in lattice models, in
any dimension. In the following work, however, we will apply it to systems whose
degrees of freedom are spins on a $2d$ lattice, and thus we will use the familiar 
'spin' terminology of statistical mechanics. 
The system is first prepared in a typical equilibrium configuration belonging to
its metastable phase between $T_{sp}$ and $T_c$. All the spins 
outside a certain region (typically spherical) of size $R$ are then {\it frozen}.
The frozen spins produce a pinning field on the border of the free sphere, which
bias the thermodynamics of the sphere. With this setup, the 'bubble' becomes a 
sort of laboratory for studying nucleation in action: it can either remain in the 
metastable phase, with free energy density $f_m$, or flip to the stable phase
with free energy density $f_s$, at the price of paying a surface energy due to the
frozen pinning field. We can thus write the partition function 
of the subsystem as the sum of two contributions,
\begin{eqnarray}
Z(R) =   \exp \left(-\beta V_d R^d f_m \right)+\nonumber\\
\label{Z}
+\exp\left(-\beta V_dR^df_s  -\beta S_dR^{d-1}\sigma\right)=\\
=Z_0\exp\left(-\beta V_dR^df_m-\beta S_dR^{d-1}\sigma\right)\nonumber \ ,
\end{eqnarray}
where,
\begin{equation}
Z_0=\exp(\beta V_dR^d \delta f)+\exp(\beta S_dR^{d-1}\sigma )\ ,
\label{z0}
\end{equation}
and $\delta f= f_m - f_s$, is the bulk free energy density difference between the two phases.
From the partition function we have that the probability for the bubble to be 
in the metastable phase is,
\begin{equation}
P_m(R)= \frac{1}{Z_0}\exp\left(\beta S_d R^{d-1}\sigma\right),
\label{Pm}
\end{equation}
while the probability of finding it in the stable phase is, 
\begin{equation}
P_s(R)= \frac{1}{Z_0}\exp\left(\beta V_d R^d\delta f \right) \ .
\label{Ps}
\end{equation}
Due to the nucleation trade-off explained above, for small $R$ the
bubble stays in the metastable phase, pinned there by the frozen
field, whereas for large $R$ it is thermodynamically more favourable
to be in the stable phase. The sharp transition between the two states
occurs when $R$ is equal to to critical nucleus size $R_c$, which is
thus defined by the equation,
\begin{equation}
P_m(R_c)=P_s(R_c) \ .
\end{equation}
The value of the critical nucleus is equal to,
\begin{equation}
R_c=\frac{\sigma}{\delta f}\ \frac{S_d}{V_d}\ .
\end{equation}
The FPF method consists in measuring the critical nucleus
$R_c$ as the point where the two probabilities are the same. The bulk
free energy density difference $\delta f$ is normally quite easy to
obtain (either numerically, or by an extrapolation of the free energy
 of the stable phase), and the numerical constants $S_d$ and
$V_d$ are known from the outset, since they depend on the chosen shape
of the bubble. As a result, the surface tension $\sigma$ becomes
available. Note that the critical nucleus size $R_c$ is up to a numerical constant
equal to the value $\widehat R$ that maximizes Eq.(\ref{DF}).

A possible objection is that this is just the same as considering the
{\it entire} system and look for stable nuclei. After the discussion
above, it should be clear that this is not the case. By considering 
the entire system, the surface tension can be obtained only on the basis of {\it dynamical} 
observations. The FPF method, instead, extracts information on the surface tension 
from {\it static} probabilities, without the need of any hypothesis 
on the unknown prefactor of the nucleation time.

We end this Section by noting that Gibbs' nucleation theory~\cite{gibbs} 
is based on the assumption of sharp edges (or thin walls) separating two homogeneous phases.
Nevertheless the theory is still valid as far as the critical nucleus is 
big enough to contain a homogeneous core. 
Several theories~\cite{ch1, ch2, fiskwidom, sarkiesfrankel} have been introduces to treat the case 
in which the nucleus is completely non-homogeneous,
defining a local free-energy in a non-uniform system. 
The FPF method uses the simple scheme of nucleation theory, but, as we shall see in the following, 
it is also able to find out whether we really are in the validity range of this 
scheme, whether a homogeneous core exists, and, in the case where the edge is not sharp, 
to study the profile of the order parameter across the boundary.

\section{\label{t>tc} The paramagnetic Ising model}

The first system we consider is the Ising model in two dimensions,
\begin{equation}
H=-J\sum_{\langle i,j\rangle} s_i s_j \ ,
\end{equation}
with $s_i=\pm 1$ and the sum is over nearest-neighbours.  Even though
our main interest is to test the FPF method within the first-order
transition induced by the presence of a magnetic field below $T_c$, we
will first study the method above $T_c$, in the paramagnetic
phase. This will be a useful warm-up exercise to get familiar with
all features of the method not strictly related to the presence of a
nucleation-driven transition. In the following we will take $J=1$ with
no loss of generality.

The main point of the FPF method is to control whether the free bubble
changes phase or not. To this aim a convenient tool is to use the
overlap $Q$ between two different configurations of the bubble, let us call
them $\{\rho_i\}$ e $\{\tau_i\}$,
\begin{equation}
Q\equiv \frac{1}{n(R)}\sum_{i=1}^{n(R)} \rho_i \tau_i = \frac{1}{n(R)}\sum_{i=1}^{n(R)} q_i
\label{Q}
\end{equation}
where $n(R)$ indicates the number of spins in the bubble, $R$
represents the bubble's size, and $q_i\equiv \rho_i \tau_i$.  In the
paramagnetic phase, there is no possibility for the bubble to flip to
another states, and both $\{\rho_i\}$ e $\{\tau_i\}$ will belong to
the paramagnet. In $d=2$, we can rewrite $Q$ as,
\begin{equation}
Q(R)=\frac{2}{R^2}\int_0^R dr\; r \, q(r)\ ,
\label{ottone}
\end{equation}
where $q(r)$ is the radial overlap, i.e. the average overlap of a
shell of spins of radius $r$ within the bubble.  In presence of a
single paramagnetic phase, it seems reasonable to assume that the
radial overlap $q(r)$ decays exponentially off the border with
characteristic length $\lambda$,
\begin{equation}
q(r)=q_0\exp\left(-\frac{R-r}{\lambda}\right) \ ,
\label{qth}
\end{equation}
where $q_0$ is the value of the local overlap at the border, which
varies with the temperature $T$. By plugging this form into
(\ref{ottone}) we get,
\begin{equation}
Q(R)=2 q_0\frac{\lambda}{R}\left[1-\frac{\lambda}{R}\left(1-\exp\left(-\frac{R}{\lambda}\right)\right)\right] \ .
\label{Qth}
\end{equation}
The length $\lambda$ gives an estimate of the thickness of the layer
directly influenced by the frozen pinning field at the border. 

\begin{figure}[t]
\includegraphics[width=12.7cm]{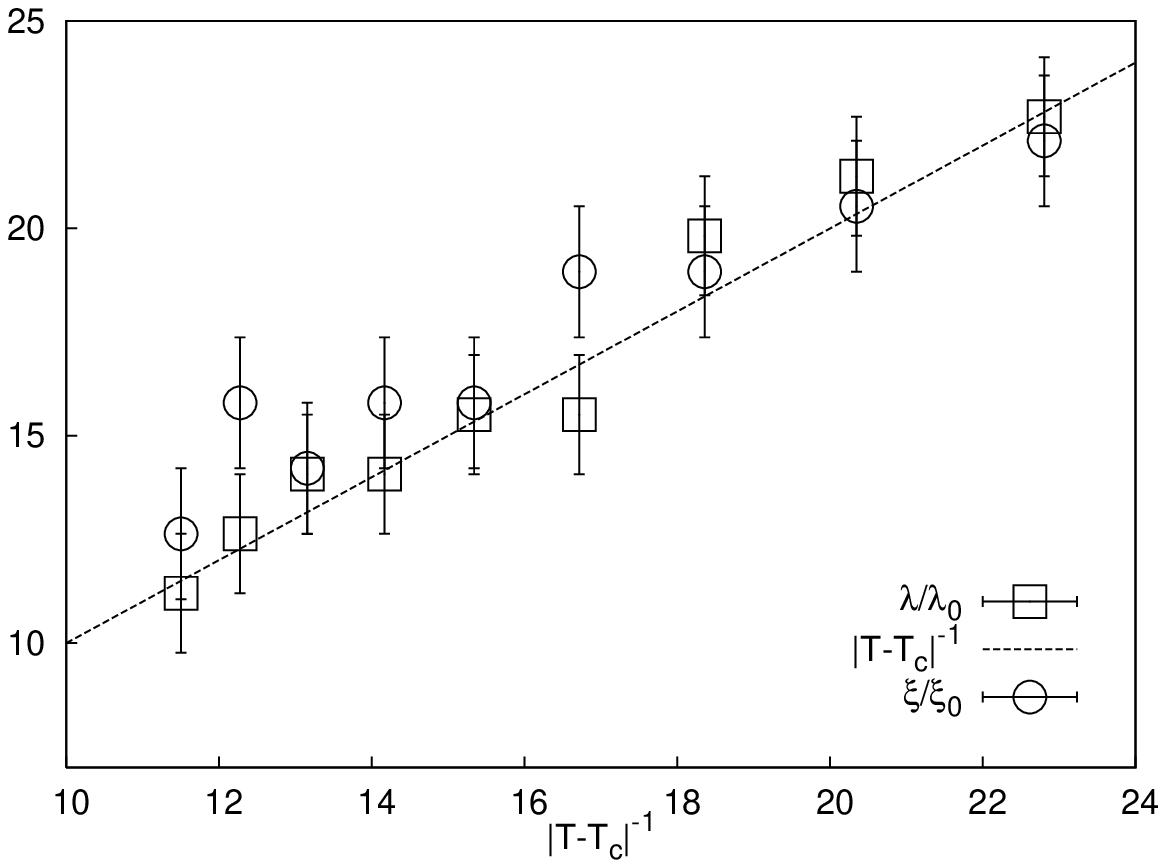}
\caption{The length-scale $\lambda$ and the correlation length $\xi$ vs
the temperature. Both lengths are normalized by their numerical prefactor in a
$|T-T_c|^{-1}$ fit.
}
\label{lambda}
\end{figure}

We performed Montecarlo simulations to study a bidimensional 
Ising square lattice ($1000\verb+x+1000$) at several temperatures in
the interval $T\in [1.015 T_c:1.039 T_c]$, and at zero magnetic field.
We freeze the system in an equilibrium configuration, and follow the evolution
of a free bubble (a disc, in fact) of radius $R$.
For every size $R$, we averaged over $200$ independent pinning fields, and over $1000$ 
Montecarlo sweeps for each pinning field.
The function (\ref{Qth}) provides a good fit to the $Q(R)$
values at the several temperatures considered.
By using standard methods we have also computed the correlation
length $\xi$ of the system, which is known to diverge as 
$(T-T_c)^{-1}$ in two dimensions. Both $\lambda$ and $\xi$ are
plotted in Figure \ref{lambda}: one clearly sees that they are
basically the same length-scale.Not surprisingly, the thickness
of the superficial layer most influenced by the pinning field 
is the same as the correlation length.

\section{\label{critclsize} The Ising model in a field}

The Ising models below $T_c$ has a first-order phase transition ruled
by the magnetic field $h$, and a transition point at $h_c=0$. At
negative values of the field the 'up' phase is metastable, and it does
eventually decay to the stable 'down' phase.  We want to use the FPF
method to evaluate the size of the critical nucleus and the value of
the surface tension. In order to have a well-defined metastable phase
we have to stay away from the kinetic spinodal, that is we have to 
choose a field whose absolute value is {\it
smaller} than the kinetic spinodal $h_{sp}(T)$ at that particular
temperature. On the other hand, the field cannot be too small:  
the free energy difference $\delta f$ which enter in eq.(\ref{tempo}) is proportional to the
field, so that when $|h|$ decreases, the nucleation time increases
exponentially. Thus, when the field is too small, a fair sampling of the
bubble's phase space, obtained by letting it flip between the two phases a
sufficiently large number of times, becomes impossible.  If we call
$h^\star$ this minimal value of the field, we have to work in the
window $h^\star <|h| < h_{sp}$. We study the FPF method at temperature
$T=0.56 T_c$, where we have found that the field interval $h\in [-0.35
:-0.23]$ satisfies these requirements.

\begin{figure}[t]
\includegraphics[width=12.7cm]{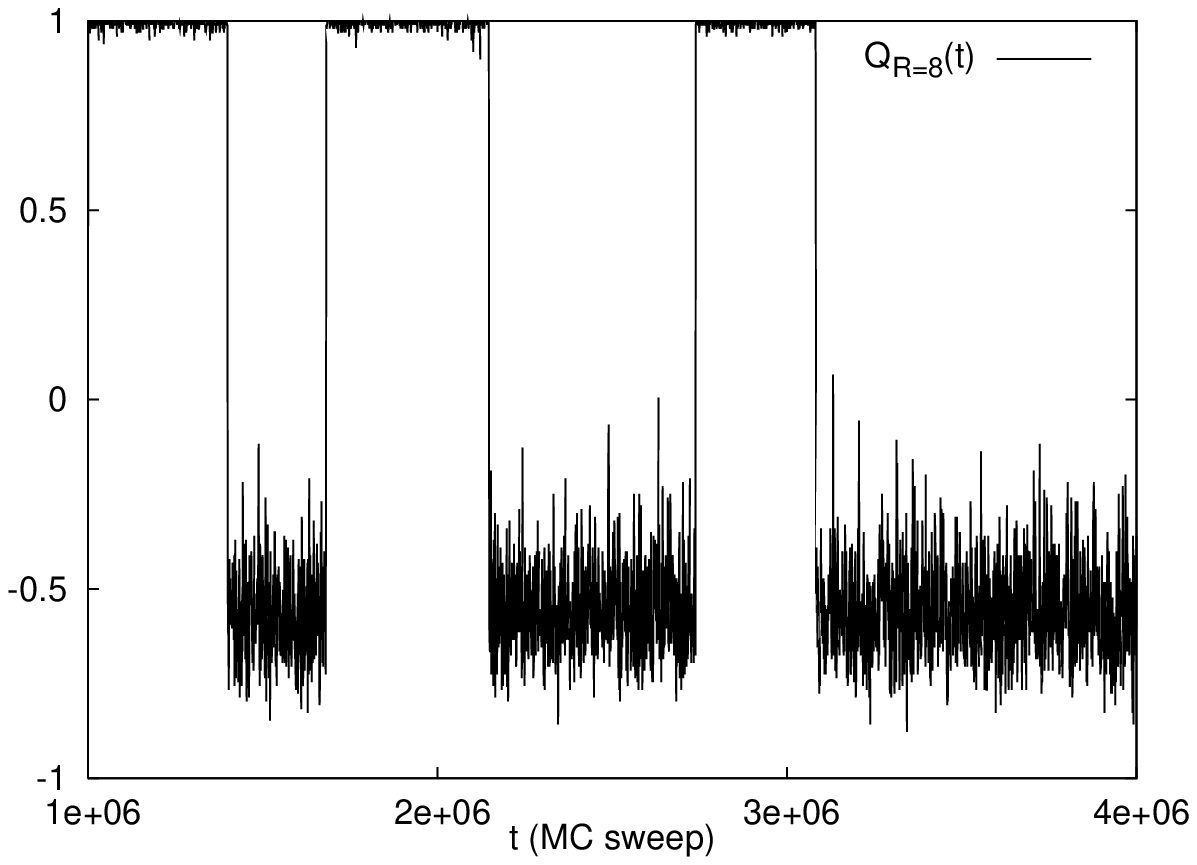}
\caption{Time evolution of the overlap of the bubble with $R=8$ 
between the initial 
metastable configuration and the configuration at time $t$.}
\label{qt}
\end{figure}

\begin{figure}[t]
\includegraphics[width=12.7cm]{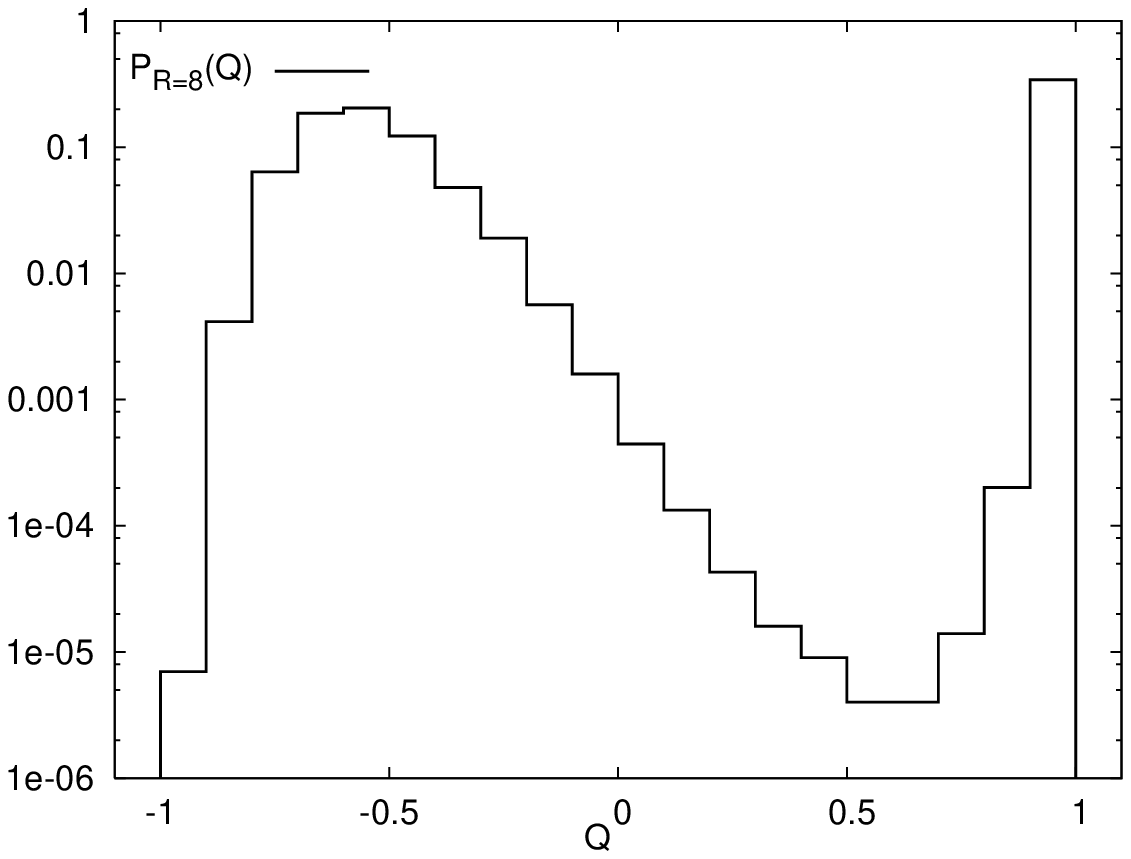}
\caption{The probability distribution of the overlap for a bubble 
with $R=8$.}
\label{Pq}
\end{figure}

In Figure \ref{qt} we see the time evolution of the overlap between the initial
metastable configuration of the bubble and its configuration at time $t$.
The size $R$ in this particular run was very close to the critical one, and
a clear flip-flop between the stable and metastable phases can be seen.
In Figure \ref{Pq} we report the probability distribution of the 
overlap.

\begin{figure}[t]
\includegraphics[width=12.7cm]{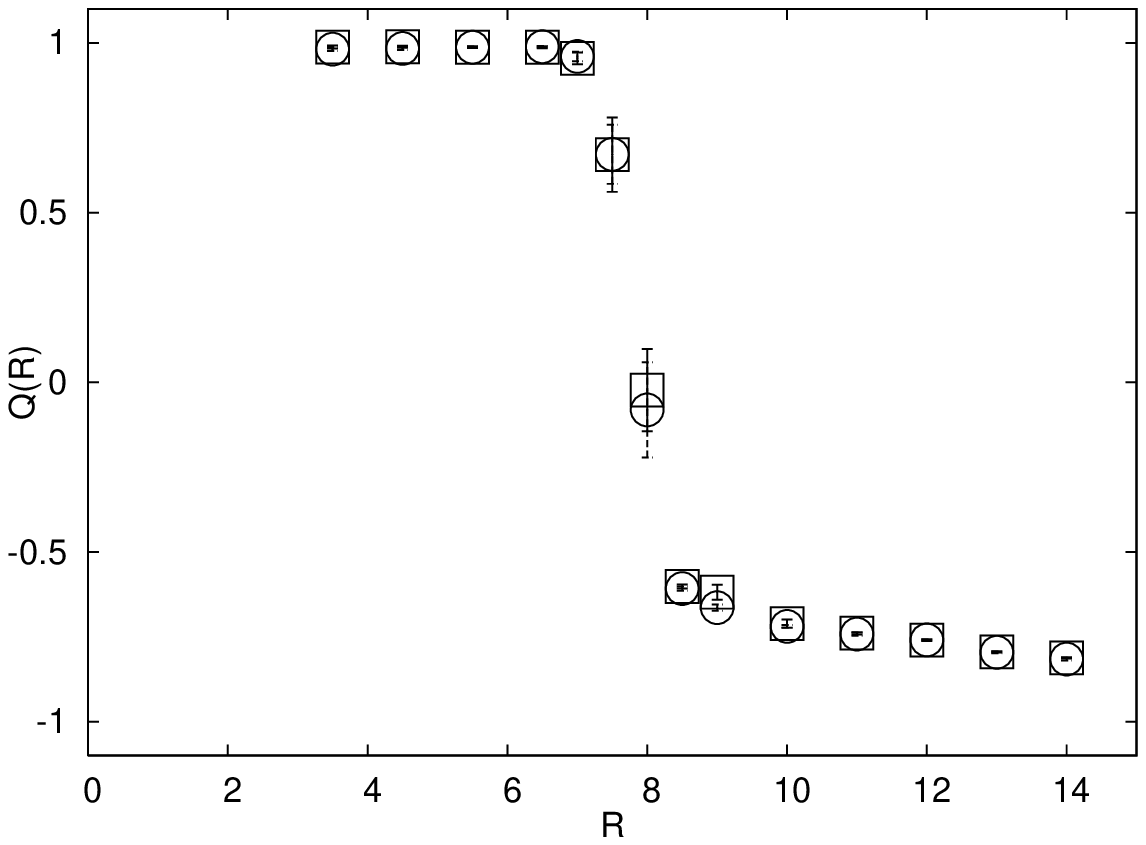}
\caption{Global overlap as a function of the size of the 
subsystems, at the temperature $T=0.56T_c$ and field $h=-0.25J$. Different symbols refer
to initializing the bubble in the stable and metastable phase, to show that 
our result derive from a fair sampling of the bubble phase space.}
\label{QhRc}
\end{figure}

The average overlap $Q$ as a function of $R$ is reported in Figure \ref{QhRc}: 
the sharp jump corresponding to the size of the critical nucleus $R_c$ is
very clear.  In order to be sure that we have sampled fairly the bubble's phase
space we initialize it both in the metastable and in the stable phase,
and check that we get the same asymptotic results (see Fig. \ref{QhRc}).
This means that we correctly thermalized the bubble.
From the Eqs.(\ref{Pm}) and (\ref{Ps}) we can write
\begin{equation}
Q(R)=P_m(R) q_{mm}(R) +P_s(R) q_{ms}(R) \ ,
\label{qab}
\end{equation}
where $q_{mm}$ and $q_{ms}$ are respectively the overlap of the
metastable phase with itself (self-overlap), and with the stable
one. These values, however, do not refer to the entire system,
but are restricted to the bubble of radius $R$. We already know from
the paramagnetic case that even with one single phase the average overlap $Q(R)$
does not decay sharply to its asymptotic value, but it rather has a slow
power-low behaviour dependent on $R$. This is due to the presence of a layer
influenced by the pinning field. For this reason we indicate a $R$
dependence into $q_{mm}$ and $q_{ms}$, which is particularly evident
after the jump if Fig.\ref{QhRc}: the overlap does not directly reach the asymptotic
value between the metastable and stable phase ($Q~-1$). Rather, the
sharp drop-off at $R_c$ becomes a smooth decay for $R>R_c$.
As we have said, this behaviour is the result of the finite interfacial 
thickness of the region between the two phases.
As in the paramagnetic phase of the Ising model the presence of a 
transition region moves the mean global overlap value 
away from the value of a uniform subsystem.

Unlike in the paramagnetic case, we now deal with the contact between two different phases,
and one may therefore think that the interfacial thickness $\delta$ has an entirely 
different origin compared to the length $\lambda$ we obtained in the paramagnet.
However, within Van der Waals theory~\cite{vanderwaals} of liquid-vapour coexistence, 
it has been demonstrated that $\delta$ diverges  
at the critical point with the same critical exponent as the 
bulk correlation length $\xi$, as we found in the paramagnetic phase. 
Using eq.(\ref{Qth}), with $\delta$ in place of $\lambda$, 
one obtains that at large $R$ the overlap $Q(R)$,
should decay as $Q(R)\sim \delta/R$.
A careful study of our data for $Q(R)$, however, shows that it decays slightly 
faster than $1/R$. This suggests that also the width $\delta$ varies with $R$,
decreasing with the surface curvature, and approaching an asymptotic
value $\delta_{\infty}$ for very large $R$ (flat surface).

The critical nucleus size can be extracted from the probability distribution
of the overlap, as the point where $P_m=P_s$, that
is the size of the bubble spending half of its time in the metastable
phase and half in the stable one.  
In particular, we can interpolate the value of $R_c$ (and calculate
its error) from the two data immediately smaller and larger than the value $P(R)=1/2$.
The values of the critical nucleus obtained at $T=0.56 T_c$ are listed in 
Tab.\ref{tabRc}.

\begin{table}[h]
\begin{tabular}{c|c|c|c}
$h$ & $R_c \pm \Delta_{R_c}$ & $\sigma \pm \Delta_{\sigma}$ & $ \sigma \cdot\pi/4$ \\
\hline
0.23  &			8.50	$\pm$	0.06	&	1.96$\pm$	0.01 &	1.54 \\
0.25  &			7.83	$\pm$	0.06	&	1.96$\pm$	0.02 &	1.54 \\
0.27  &			7.27	$\pm$	0.01	&	1.963$\pm$	0.003 &	1.542\\
0.29  &			6.87	$\pm$	0.04	&	1.99$\pm$	0.01 &	1.564\\
0.31  &			6.28	$\pm$	0.01	&	1.945$\pm$	0.004 &	1.528\\
0.33  &			6.2	$\pm$	0.1	&	2.05$\pm$	0.04 &	1.61\\
0.35  &			5.51	$\pm$	0.01	&	1.928$\pm$	0.005& 1.514\\
\hline
\end{tabular}
\caption{The critical sizes $R_c$, the corresponding surface tension values 
$\sigma$ at several magnetic
field values $h$ and temperature $T=0.56T_c$.}
\label{tabRc}
\end{table}

\begin{figure}[t]
\includegraphics[width=12.7cm]{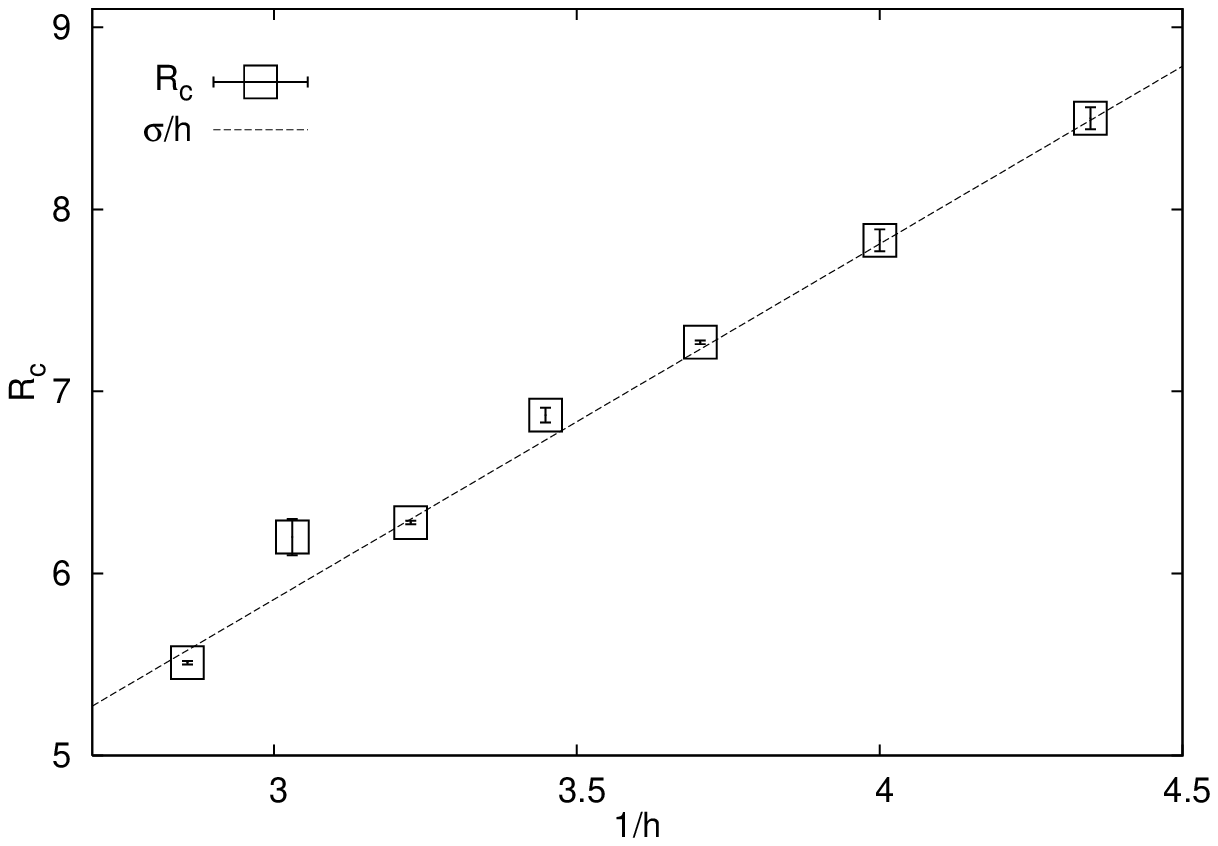}
\caption{Linear behaviour of the  critical size $R_c$ with 
respect to the inverse of the magnetic field. The slope of this curve is 
the surface tension $\sigma$.}
\label{QhRc1}
\end{figure}

From classic nucleation theory, for a disc in two dimensions, we have,
\begin{equation}
R_c=\frac{2\sigma}{\delta f}.
\label{Rc2d}
\end{equation}
At our values of temperature and field one can prove that a very good approximation is~\cite{TESI},
\begin{equation}
\delta f \sim 2h\ ,
\end{equation}
thus giving,
\begin{equation}
\sigma = R_c \, h \ ,
\label{sarponella}
\end{equation}
giving the values of surface tension $\sigma$ listed in the third
column of Tab.1.
As expected, the surface tension $\sigma$ very weakly depends 
on $h$. In fact, if we plot the critical nucleus size against the 
inverse of the field (Fig.\ref{QhRc1}), we find a very good linear 
behaviour. A linear fit gives $\sigma=1.94$, compatible with the 
values in Tab.1.

We now want to compare the FPF estimate of the surface tension with the 
exact result of Onsager~\cite{1944PhRv...65..117O}: the expression 
of the surface tension for an interface parallel to the lattice axes
in the two-dimensional Ising model at zero field is, 
\begin{equation}
\sigma_{\rm ONS} =\left\{2J-k_BT\log[\coth(J/k_BT)]\right\} \ .
\label{s_onsager}
\end{equation}
At our working temperature this relation gives,
\begin{equation}
\sigma_{\rm ONS}=1.467 \ .
\end{equation}
Before we can compare with our result, we have to make to remarks.
First, Onsager's value holds at $h=0$, whereas we are working at non-zero field, 
therefore we do not expect the two values to be identical. However, as
already said, the dependence of $\sigma$ on the field is weak: only close
to the spinodal the surface tension is expected to abruptly decrease.
Secondly, and more importantly, we computed the surface tension for a 
disc, whereas Onsager's value refers to an interface parallel to the 
(square) lattice. In this latter case, the number of bonds is equal to
the number of spins along the surface, whereas in the disc the average
number of spins is equal to $2\pi R/l$ and the number of bonds is
$8R/l$ ($l$ is the lattice spacing). Therefore the surface tension for 
the straight surface must be smaller than that of the disc by
a factor equal to the ratio between spins and bonds, i.e. $\pi/4$.
These values are listed in the fourth column of Tab.1. 

The agreement
between our estimate and Onsager's value is on average within 5\%.
Considered the remarks above about the fact that we are at 
non-zero value of the magnetic field, this result is encouraging. 
Note that to get this result we had to 
perform no infinite size extrapolation, and the computational time
is therefore relatively limited. However, if we wanted to obtain 
the value of $\sigma$ for $h\to 0$, in order to exaclty check
Onsager's results, we would get larger and larger critical sizes
$R_c$, and this {\it would} in fact require an infinite size
limit. Considered that, as we have already said, the dependence
of the surface tension of the field is very small, it is just a
matter of convenience to perform or not this limit. On the other
hand, if one needs the surface tension at {\it finite} values
of the field, then the value of $R_c$ is finite, and the 
large $N$ limit is not at all necessary.

To conclude this Section, we study the dependence of the
critical nucleus size $R_c$ on the temperature, at fixed magnetic 
field. We use the FPF method with 
magnetic field $h=-0.25$ and temperatures $0.56 T_c$ and $0.64 T_c$.
The resulting values of the of the critical size at the listed temperature are,
$R_c(0.56T_c)=7.83\pm0.06$ and $R_c(0.64T_c)=7.24\pm0.01$.
The increase of the tempertaure induces a slight decrease 
of the critical size $R_c$. Why is that?
When the temperature is increased, we have a decrease of both 
$\sigma$ and $\delta f$. Yet, $\sigma$ variation with $T$ is almost linear 
according to Onsager equation, while at temperatures well below $T_c$ 
and weak magnetic field, the change of $\delta f$ is weak, since it
is mainly given by the change in the magnetization $m$, which saturates
rapidly to $1$, far from $T_c$.
Therefore the decrease of $\sigma$ is sharper than that of $\delta f$, 
and for this reason $R_c=\sigma/\delta f$ decreases when $T$ is increased
far from $T_c$. 

From the value of $R_c$ at $T=0.64 T_c$
and from the corresponding $\delta f$, we get,
\begin{equation}
\sigma=1.810\pm0.003 
\end{equation}
and 
\begin{equation}
\sigma \cdot \pi/4 =1.421 \ .
\end{equation}
Onsager's surface tension 
at the same temperature is,
\begin{equation}
\sigma_{\rm ONS}=1.29 \ .
\end{equation}
At this temperature the agreement between the FPF surface tension and Onsager's one is 
worse than the one at $T=0.56T_c$, even though still within 10\%.

We note that in all cases considered the FPF method gives a surface tension
slightly larger than the exact zero-field result. Although the field 
is non-zero in our case, we believe the discrepancy mainly comes from a different effect, 
namely the fact that our method overestimates the size of the critical radius $R_c$ 
by an amount roughly equal to the interfacial thickness. The reason for this is
quite simple: due to the frozen nature of the external environment, the order parameter
can change its value only {\it within} the disc, and not in going {\it through} the disc.
The intermediate value (zero, for the magnetization), which defines the {\it bona fide} 
border of the critical droplet, is therefore attained at the interior of the disc,
and it is thus smaller than the one we measure, by an amount approximately equal to 
half of the interfacial thickness. Indeed, as we have seen, 
the discrepancy is larger at larger temperature,
where one expects to have a thicker interface. Of course, such effect is negligible when the 
size $R_c$ of the bubble is much bigger than the interfacial thickness.

\section{\label{ctls} A model with temperature-driven first-order transition}

The parameter tuning the first-order transition in the Ising model is
the magnetic field, and both the stable and the metastable phases are
ordered (ferromagnetic). Due to this, the frozen pinning field
produced by the metastable configuration on the border of the bubble
is quite similar to a standard magnetic field (up to thermal
fluctuations), and this makes the setup of the FPF method somewhat
predictable. In this section we will apply the FPF method to a model
with a first-order transition driven by temperature, rather than
magnetic field. In such a model we will consider a disordered
(liquid-like) metastable phase, in contact with an ordered
(crystal-like) stable phase. In this way we will test the FPF method
in a far less trivial context than the Ising model.

\begin{figure}[t]
\includegraphics[width=8. cm]{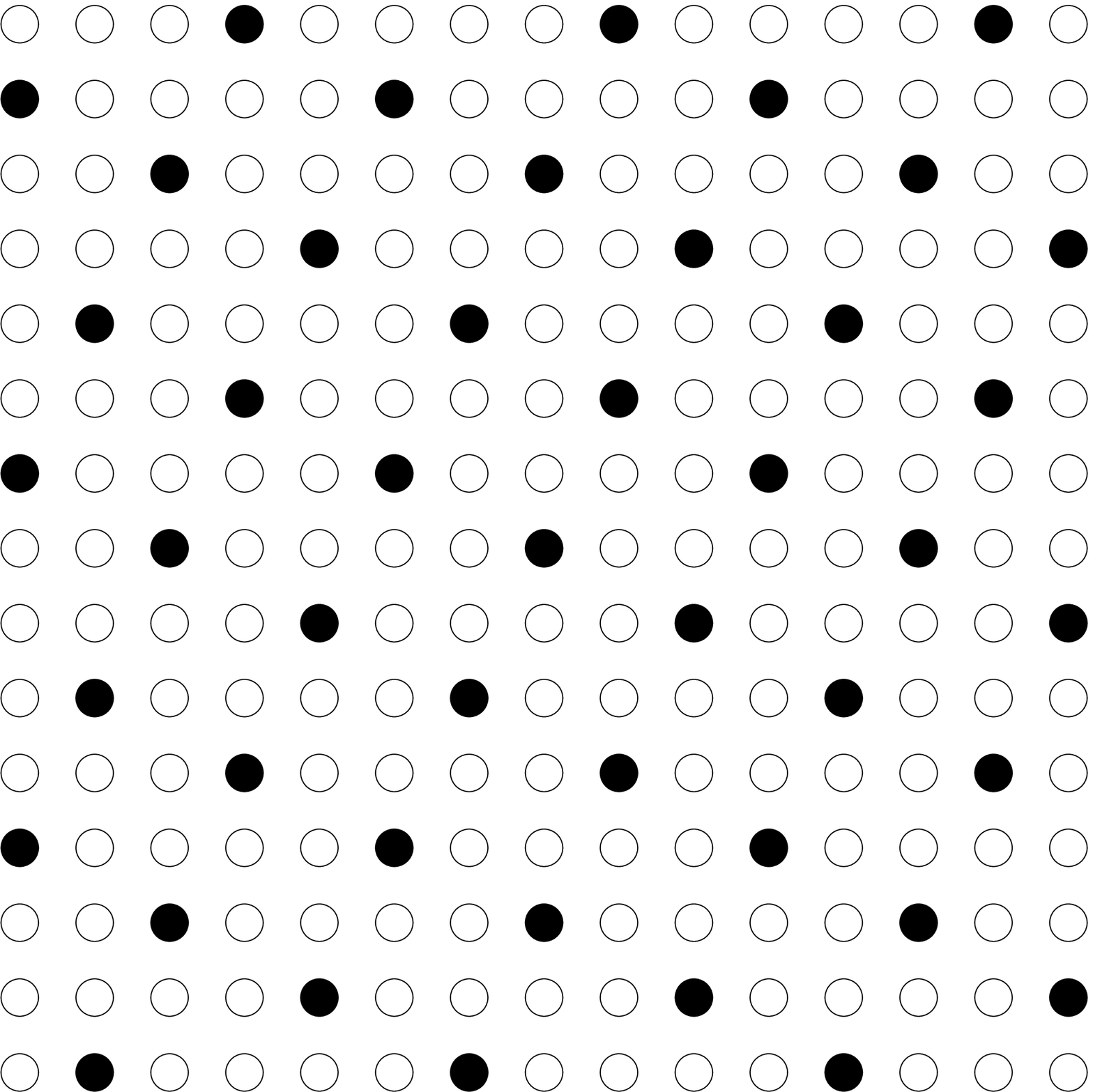}
\caption{Crystalline ground-state of the CTLS model.}
\label{ctlsfond}
\end{figure}

The two-dimensional lattice model we consider is called
CTLS (Coupled Two Levels System), and it was studied
in~\cite{grigera2002, 2003JChPh.118.6974C}. The Hamiltonian of the model is,
\begin{equation}
H=J\sum_{i=1}^N f_i(1+\sigma_i)
\label{HCTLS}
\end{equation}
where $N$ is the number of spins and $f_i$
indicates the product of the $4$ nearest neighbour spins, 
$$f_i=\sigma_i^N\sigma_i^S\sigma_i^E\sigma_i^O \ ,$$
where $N$ stands for north, $S$ for south, and so on.
The CTLS has a crystalline ground-state (Fig.\ref{ctlsfond}), and the 
low temperature crystalline phase is separated by a high temperature
disordered `liquid' phase by a first-order transition at 
a melting temperature $T_m=1.30 J/k_B$. The multibody interaction 
in the Hamiltonian introduces frustration in the system, and due
to this the CTLS exhibits a behaviour typical of supercooled and
glassy systems~\cite{2003JChPh.118.6974C}, with many disordered local 
minima of the energy. Below the melting temperature, the liquid enters a
metastable region, which is bounded by a kinetic spinodal at a 
temperature $T_{sp}$:
below $T_{sp}$ the relaxation time of the metastable liquid phase 
becomes larger than the crystal nucleation time. The metastable phase
becomes therefore kinetically unstable below $T_{sp}$. Our test
of the FPF method will thus be performed for
$T_{sp} < T < T_m$.

\begin{figure}[t]
\includegraphics[width=12.7cm]{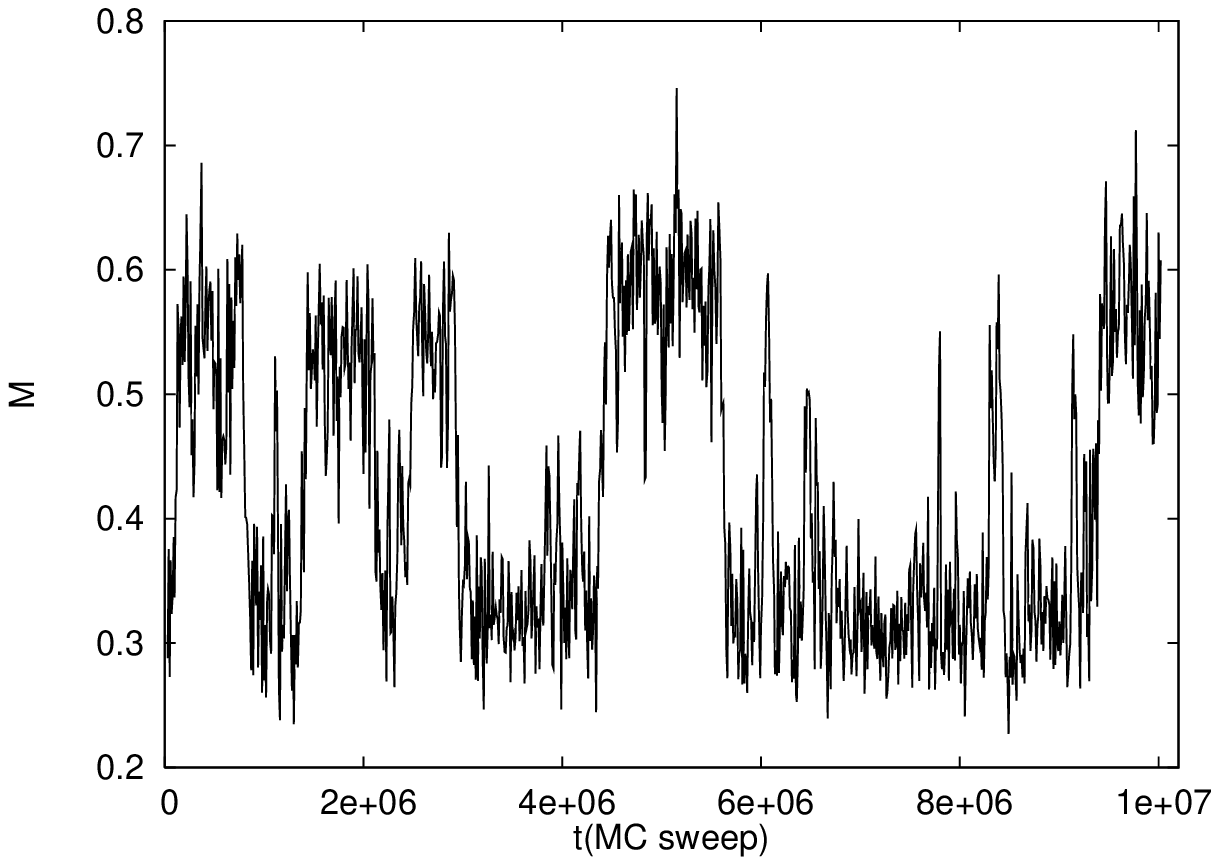}
\caption{The average crystal mass of a disc with $R=29$ as a function of time.}
\label{mcrt}
\end{figure}

\begin{figure}[t]
\includegraphics[width=12.7cm]{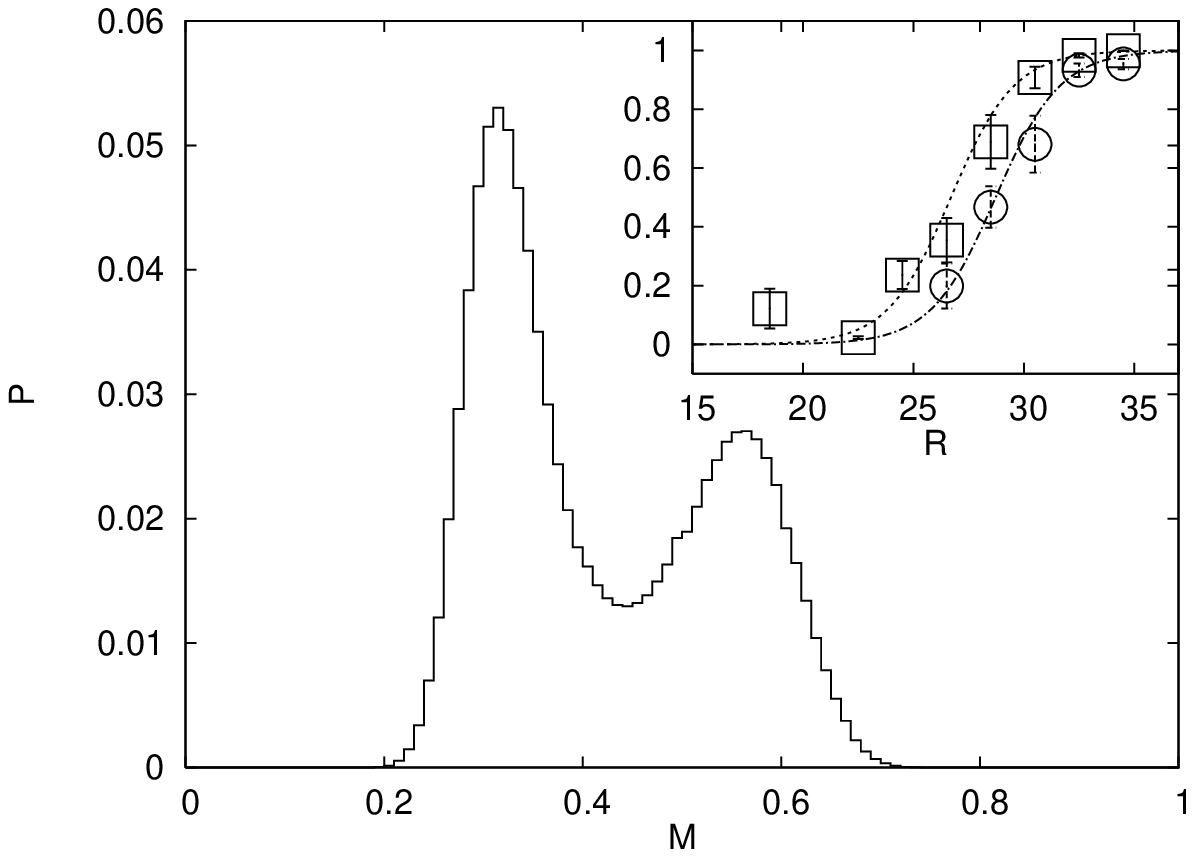}
\caption{Probability distribution of the crystalline mass of a disc with $R=29$.
The inset shows the probability $P_{c}$ as a function of the size of the 
subsystems at the temperature $T=0.931 T_m$ (the squares) and $T=0.938 T_m$ (the circles).}
\label{Pmcr}
\end{figure}

We equilibrate the system in a metastable liquid configuration below $T_m$,
and we freeze all spins outside a disc of radius $R$. In order to check
what phase the disc is in, we must introduce a suitable order parameter.
Following~\cite{2003JChPh.118.6974C} we define the local crystalline
mass $m_i\in [0:1]$. In the temperature range we will consider we have
$\langle m\rangle \sim 1$ in the stable crystalline phase, and $\langle m \rangle
\sim 0.3$ in the metastable liquid phase. Therefore, the crystalline mass
is a good order parameter to distinguish the liquid from the crystal phase.
The average crystal mass of the disc,
\begin{equation}
M=\frac{1}{n(R)}\sum_{i=1}^{n(R)} m_i \ ,
\end{equation}
where $n(R)$ is the number of spins within the disc, is plotted as a
function of time for a particular value of $R$ in Figure \ref{mcrt}. This value
of $R$ is quite close to the critical nucleus. Even though the jumps
between stable and metastable phases are less clear-cut than in the
Ising case, the histogram still shows a very evident bimodal
distribution of the order parameter (Figure \ref{Pmcr}). The two peaks
correspond to the probabilities of the liquid (LQ) and crystalline
(CR) phases. By changing the radius $R$ of the bubble the balance
between the two phases changes according to nucleation's prediction.
The critical nucleus size $R_c$ corresponds to the point where the two
peaks have the same weight. In the inset of Figure \ref{Pmcr} we plot the probability
$P_{CR}$, i.e. the weight of the crystalline peak, as a function of
$R$.
As we did in the Ising model 
we can now obtain the value of the critical nucleus: 
$R_c=27.4\pm0.4$ at $T=0.931 T_m$ and 
$R_c=28.8\pm0.5$ at $T=0.938 T_m$.
Using the liquid-crystal bulk free energy difference $\delta f(T)$ calculated in \cite{2003JChPh.118.6974C}, 
it is finally possible to find the corresponding surface tension values:
$\sigma(0.931 T_m)=0.262\pm0.004$ and $\sigma(0.938 T_m)=0.250\pm0.004$. 
As in the Ising model, the surface tension is very weakly dependent on the 
parameter driving the transition (the temperature, in this case), provided 
that we are far enough from the spinodal point.

\begin{figure}[t]
\includegraphics[width=12.7cm]{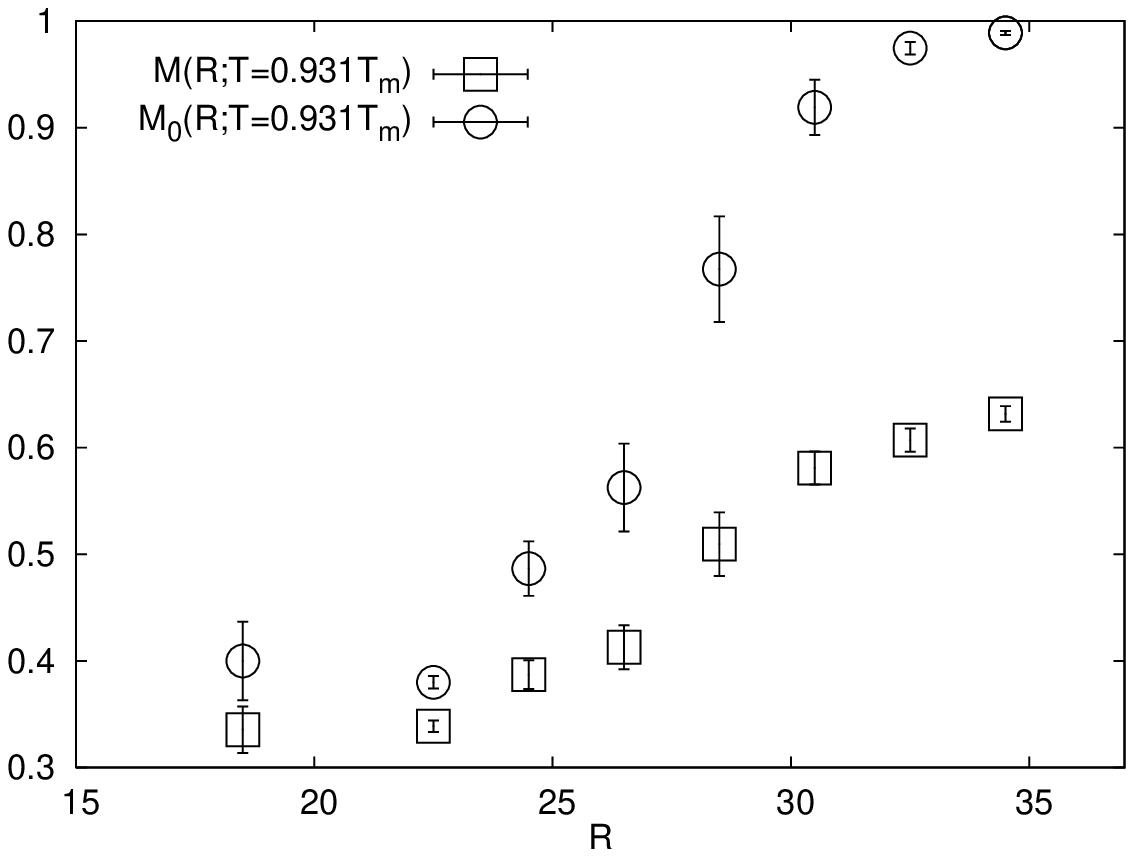}
\caption{Global crystalline mass as a function of the size of the 
subsystems at the temperature $T=0.931 T_m$ and the crystalline mass of the 
center of radius $a=9.5$ at the same temperature.}
\label{McrR}
\end{figure}

We note from the distribution of $M$ that even for $R > R_c$, the position
of the (unique) crystalline peak is quite far from its equilibrium value $M\sim 1$.
This is also clear from the average value of $M$ as a function of $R$ in 
Figure \ref{McrR}. If we assume that the 
radial order parameter $m(r)$ goes exponentially fast to $1$ off the border,
\begin{equation}
m(r) = 1 - (1-m_0)\exp(-r/\delta) \ ,
\end{equation}
then, as we have seen in Section \ref{t>tc}, the large $R$ behaviour of $M(R)$ is,
\begin{equation}
M(R) \sim 1 - \delta/R
\end{equation}
Therefore, the reason why the behaviour of $M(R)$ in Figure \ref{McrR} 
is so smooth, is because
the interfacial thickness $\delta$ is quite large in the CTLS. This suggests to
check just the center of the bubble, rather than the entire disc, in order to 
get rid of the border effects and only monitor the inner thermodynamics 
of the disc.
More precisely, we can average the order parameter over a small circle of 
{\it fixed} 
radius $a$ around the center, and study the core order parameter $M_0(R)$.
A simple calculation gives,
\begin{eqnarray}
\label{Mcr20}
M_0(R)=1- \frac{2(1-m_0)}{a^2}\int_0^a dr\; r\, e^{-\frac{R-r}{\delta}}\\
=1- \mathcal M \; e^{-\frac{R-a}{\delta}}
\end{eqnarray}
with 
\begin{equation}
\mathcal M=\frac{2\delta}{a}(1-m_0)\left\{1-\frac{\delta}{a}\left[1-e^{-\frac{a}{\delta}}\right]\right\}.
\label{Mpre-exp}
\end{equation}
At fixed $a$, $M_0(R)$ increase exponentially fast in $R$, and due to this
the jump at $R_c$ of $M_0(R)$ is much sharper than the one of $M(R)$
(see Figure \ref{McrR}). Therefore, as a way to check whether the 
bubble exposed to the frozen pinning field has switched state or not, the core order
parameter $M_0(R)$ is certainly a better choice than $M(R)$.

\begin{figure}[t]
\includegraphics[width=12.7cm]{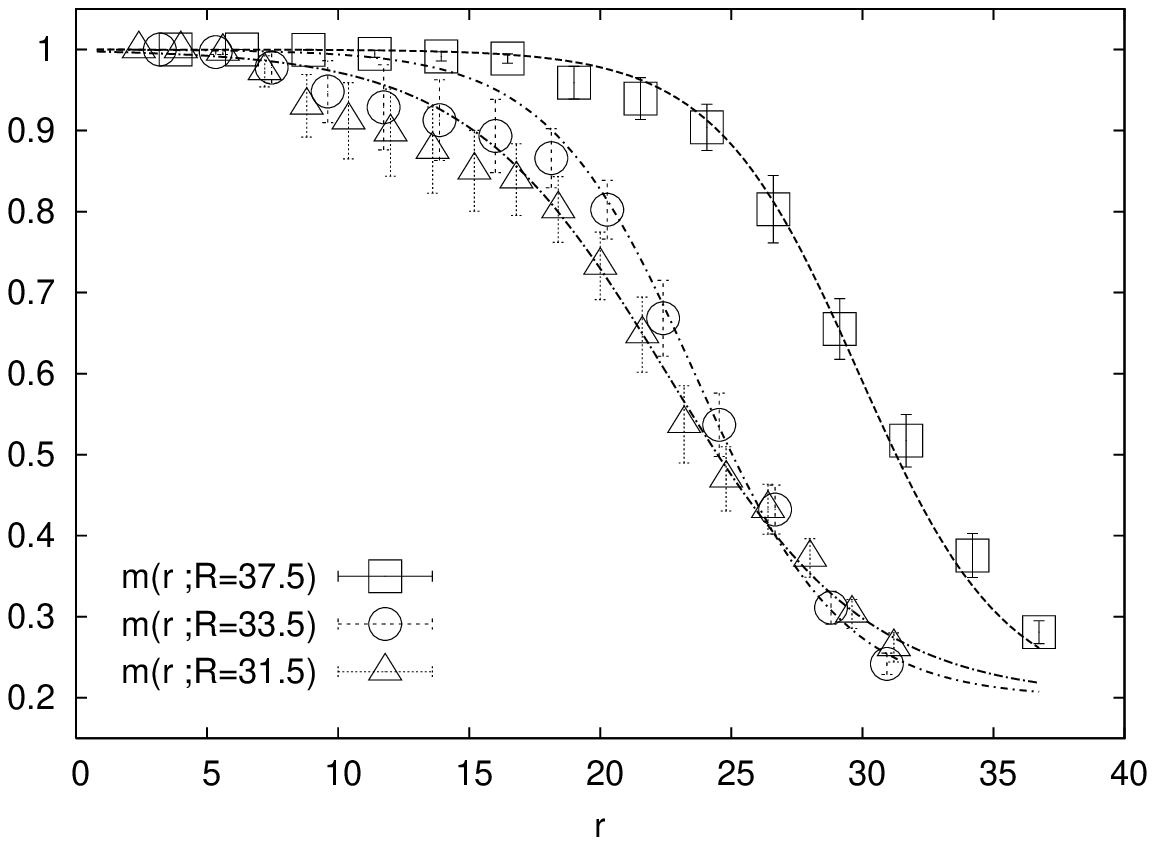}
\caption{Local crystalline mass as a function of $r$, with $R=31.5$, $R=33.5$ and $R=37.5$.}
\label{Mcrloc}
\end{figure}

According to equation (\ref{Mcr20}), $\log[(1-M_0(R))/{\mathcal M}]$ 
must be a linear function of $(a-R)/\delta$. This is very well verified by our data, 
and a linear interpolation gives an estimate of the interfacial thickness, $\delta=8.6\pm0.6$.
This shows that in the CTLS $\delta$ is much larger than in the Ising model.
It becomes therefore interesting to study the profile of
the order parameter $m(r)$ across the interface.  According to 
Cahn and Hilliard~\cite{ch1}, 
the order parameter across the liquid-vapour
interface can be described by an hyperbolic-tangent of width
$\delta$. We can make a similar fit in our case for the crystal mass,
\begin{equation}
m(r)=\frac{1}{2}\left[1+m_0 + (1-m_0) \tanh\left(-\frac{R-r}{\delta}\right)\right] \ .
\label{MinterfaceCH}
\end{equation}
This functional form works well only for large bubbles 
(large $R$, Figure \ref{Mcrloc}). The reason for this is
that for $R< R_c+2\delta$, the transition region is forced 
into a region sharper than the natural one
and it cannot take the profile that it would have in a spontaneously 
formed nucleus. 
The interfacial thickness $\delta$ from equation (\ref{MinterfaceCH}) 
is $\delta=11\pm2$, compatible with the one form eq.(\ref{Mcr20}).

\begin{figure}[t]
\includegraphics[width=12.7cm]{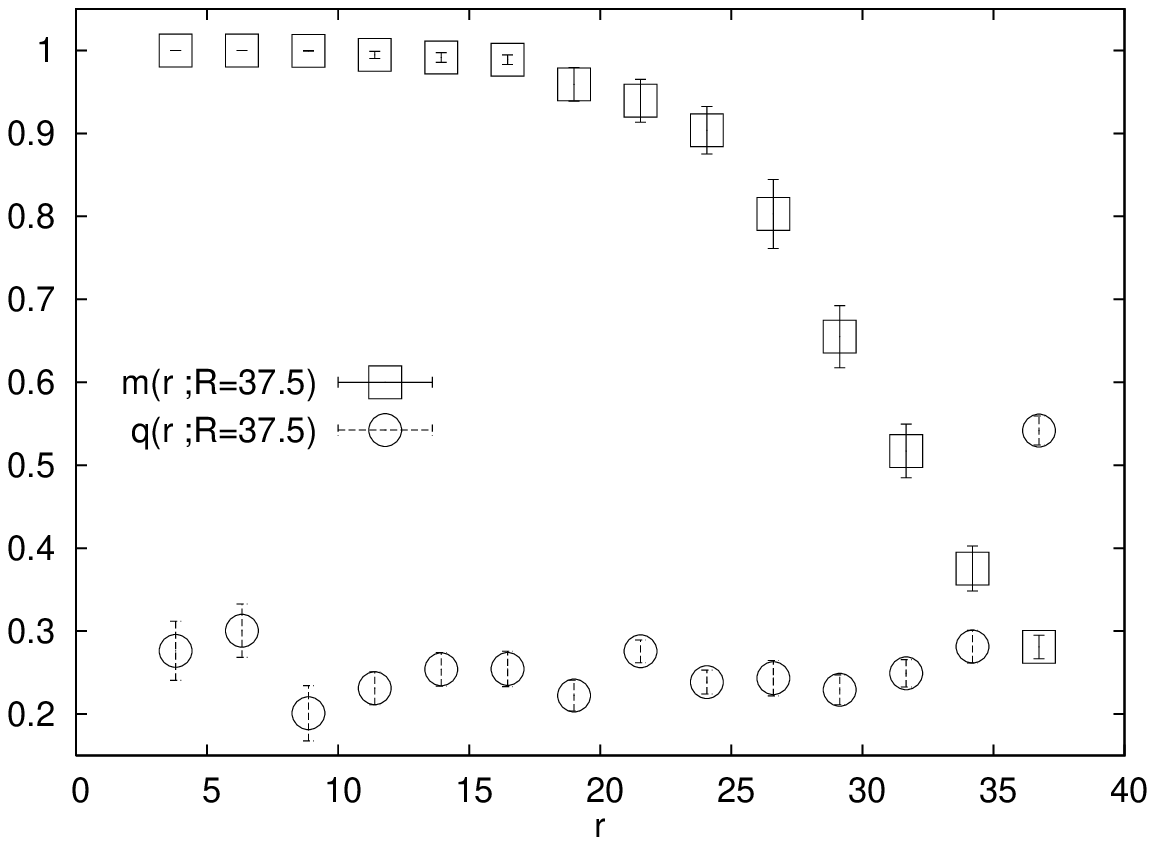}
\caption{The local crystalline mass and the radial overlap as a function of $r$,  with $R=37.5$.}
\label{Mcrqloc}
\end{figure}

Finally, we want to understand more carefully what is the effect of the frozen
pinning field produced by the metastable liquid surrounding the
bubble. From the profile of the order parameter in Figure \ref{Mcrqloc}, it is
clear that there is a wide region close to the border where the
configuration is no longer crystalline, and it is becoming liquid. The
interesting fact is that this intermediate liquid configuration is
anyway very different from the {\it initial} liquid configuration the bubble
was in. This can be seen by measuring the radial overlap $q(r)$
between the asymptotic configuration and the initial configuration of
the disc (Figure \ref{Mcrqloc}). 
The overlap is very close to zero within the disc,
telling us that the bubble has abandoned its initial liquid configuration.
What is interesting is that $q(r)\sim 0$ even in the intermediate
crossover region: this region is therefore almost liquid, but a {\it
different} liquid configuration from the initial one.  In other words,
what we see here is that the frozen pinning field is attracting the
sphere toward the liquid {\it phase}, rather than just toward the liquid {\it
configuration} in which the system was frozen. Given that two
different configurations of the same liquid phase have almost zero
overlap, then $q\sim 0$ even in the region where $m$ is rapidly decreasing. This
phenomenon was impossible to see in the Ising model, since in that
case the pinning metastable phase had a very small entropy, 
and there was little
difference between configurations belonging to the same phase.

\section{\label{caveat} The problem of the equilibrium shape}

The problem of what is the equilibrium shape (ES) of the critical
nucleus is a very old and, in general, very difficult one
\cite{wulff01}. When the system is isotropic, the surface
tension is independent of the orientation, and
it is easy to demonstrate that the ES is a sphere.
However, in non-isotropic systems the situation is not as simple as that.
Within the FPF method the droplet's shape is imposed
from the outset, and it is therefore reasonable to ask to what extent
the surface tension obtained from the FPF depends on the imposed
shape. As we are going to explain, there are in fact two distinct 
issues related to this point.

First of all, irrespective of nucleation and of the shape of the critical droplet,
there is the problem that in non-isotropic systems the surface tension depends 
on the orientation of the surface itself. This is what happens typically on 
a lattice. In the Ising case, studied above, we get different values of the 
tension when the surface is parallel to the axes, diagonal, or circular.
However, as we have seen, knowledge of the symmetry of the lattice allows
us to relate all these values one to another by simple numerical factors
(respectively $\sqrt{2}$ and $4/\pi$). This is something that can in 
principle be done for every system whose sysmmetries we have enough 
knowledge about. If, on the other hand, we have no a priori knowledge 
about interfacial orientations, the best we can do is to apply the FPF 
method using any convenient shape, and get an effective surface tension function of
the (possibly many) elementary values, $\sigma_1, \sigma_2, \dots$.

A second, distinct, problem is the one of the optimal equilibrium shape (ES) 
of the critical nucleus, also known as Wulff's shape \cite{wulff01}.
What happens if the shape imposed by the FPF method is very different
from the ES?
Imagine a system where the ES is very anisotropic, for example a 
long rectangle, with critical values of the sides $L_1 \gg L_2$, 
and assume that we use a disc of radius $R$ within the FPF method. 
One may argue that as long as $2R <L_1$ the critical (rectangular) 
droplet cannot be formed, and therefore the disc does not change state,
whereas for $R> L_1$ we will observe the formation of rectangular 
nuclei within the disc. This is, however, not the case: in the 
Ising model, where the ES is much more similar to a square than a
disc at low temperatures \cite{rottman81}, we never observe the 
formation of squares within the imposed disc (unless the disc is really
{\it much} larger than the critical nucleus, such that the disc itself
becomes an independent sub-system); on the contrary, 
when a change of state takes place, it is always the entire disc that
participates to it. 
This means that the whole FPF shape switches state whenever the 
balance betweeen area and surface {\it of that particular shape}
becomes convenient. We believe this is due to a sort of nucleation
`pressure' within the imposed shape. It is clear, however, that 
this point needs further clarification.

\section{\label{conclusioni} Conclusions}

We described a novel method for the determination of the critical
nucleus size, and of the surface tension between stable and metastable
phases in systems with first-order phase transition. The main idea of
the method is to stabilize the metastable phase of the system by
freezing it outside a bubble of radius $R$.  By studying the
thermodynamics of the bubble as a function of $R$, we unveil the
balance between the surface tension contribution, forcing the bubble
to stay in the metastable phase by means of the frozen pinning field,
and the free energy difference drive to switch to the stable phase.
The critical radius $R_c$ is defined as the point where these two 
contributions are equal. What we observe is therefore a sort of
first-order transition as a function of the parameter $R$, the 
size of the bubble.

We note that compared to other methods, the FPF method does 
not resort to a dynamical study of the nucleation events
happening spontaneously in the system. Due to this, the FPF
method avoids all the pitfalls related to the dynamical
prefactor of the nucleation time, and to finding an accurate
way to distinguish nucleation from growth. Moreover, our method
does not need a $N\to\infty$ extrapolation to produce a 
result, unless one needs to use it at the coexistence
point, where there is no free energy difference between
the two phases, and $R_c\to\infty$.

We used the FPF method to compute the surface tension in the
two-dimensional Ising model in a field, where we get a result
fairly consistent with Onsager exact calculation. Secondly, we analyzed a
model (the CTLS) where a liquid-crystal first order transition driven
by temperature exists. In this case, despite the fact that the
metastable phase (the liquid) is disordered, we still find a very
sharp bimodal probability distribution of the order parameter, and thus the
determination of the critical nucleus and of the surface tension is
very easy to do. In this model, the interface between the ordered and the
disordered phase is far from being thin, stretching the very validity
of Gibbs' nucleation theory. Moreover, we saw in a clear way that
the effect of the disordered (liquid) frozen pinning field is 
to attract the bubble to the disordered phase, and not simply 
the initial disordered configuration.

We believe the FPF method could be of some help also in other systems,
including off-lattice systems. The fact that it can only be used away
from the coexistence point should not be a major problem.  The main
aim of the method is to compute the surface tension, but this quantity
is typically very weakly dependent on the parameter tuning the
first-order transition, provided that we are not close to the spinodal
point, where anyway the metastable phase is ill-defined.

\vskip 0.5 truecm
\centerline{{\bf Acknowledgments}}

This work was directly inspired by the theoretical experiment proposed 
to test the mosaic scenario in supercooled liquids by G. Biroli and J.P. Bouchaud 
in~\cite{2004JChPh.121.7347B}. This experiment was later implemented 
numerically in a glassy system by T.S. Grigera, P. Verrochio and A.C. in~\cite{2006cond.mat..7817C}.
We thank all of them for their feedback. We also thank G. Parisi
for many important suggestions, and I. Giardina, V. Lecomte, and E. Marinari for interesting
discussions.

\bibliography{bib}
\listoffigures
\listoftables
\end{document}